\documentstyle[12pt]{article}

\newtheorem {theorem}{Theorem}

\newcounter {rem}

\newcounter {exa}
\newenvironment {example} {\refstepcounter{exa}\par\medskip\noindent{\bf
Example \arabic{exa}\,\,} }{\par}
\newcounter {def}

\newenvironment {Remark} {\par\medskip\noindent{\bf Remark\,\,}} {\par}

\newenvironment {proof} {\par\medskip\noindent{\bf Proof\,\,}}
{\qed\par}

\def \Real {I\!\!R}

\def \Complex {\,\prime\mskip-2.5\thinmuskip C}

\def \lnorm#1\rnorm {\vphantom{#1}\left\|\smash{#1}\right\|}
\def \lmod#1\rmod {\vphantom{#1}\left|\smash{#1}\right|}

\newcommand \bydef {\stackrel{\mbox{\scriptsize def}}{=}}
\newcommand \qed {\,\rule[-.23ex]{1.6ex}{1.6ex}}

\renewcommand \phi {\varphi}
\renewcommand \rho {\varrho}

\author{A~A~Balinsky\thanks{e-mail: balin@leeor.technion.ac.il} \ and
Yu~M~Burman\thanks{e-mail: mat9304@technion.technion.ac.il}\\[0.3cm]
{\small Department of Mathematics}\\
{\small Technion --- Israel institute of Technology}\\
{\small 32000 Haifa, ISRAEL.}}
\title{Quadratic Poisson brackets compatible with an algebra structure}
\date{}

\newcommand \Mat {\mathop{\rm Mat\,}}
\newcommand \Symm {\mathop{\rm Symm\,}}
\newcommand \Skew {\mathop{\rm Skew\,}}
\newcommand \SU {\mathop{\rm SU\,}}
\newcommand \Aut {\mathop{\rm Aut\,}}

\begin{document}
\maketitle
\begin{abstract}
Quadratic Poisson brackets on a vector space equipped with a bilinear
multiplication are studied. A notion of a bracket compatible with the
multiplication is introduced and an effective criterion of such
compatibility is given. Among compatible brackets, a subclass of
coboundary brackets is described, and such brackets are enumerated in a
number of examples.
\end{abstract}

\subsection*{Introduction}
Investigation of many specific physical systems leads to a problem of
quantization of Poisson brackets on spaces equipped with additional
structures.
Quantum algebras, or q-deformed algebras, have been
useful in the investigation of many  physical problems.
The most studied case is when this
structure in a Lie group structure, this leads to a celebrated quantum
groups theory (see \cite{D}).
As a matter of fact, the research in q-groups was indeed
originated from physical problems.
The interest in q-groups and q-deformed algebras
arose almost simultaneously in
statistical mechanics as well as in conformal field theories,
in solid state physics as well as in the study of topologically
non-trivial solutions of nonlinear equations.

As usual, Poisson bracket $\{\cdot,\cdot\}$ is understood as a Lie algebra
structure on the space of smooth functions $C^\infty(M)$ satisfying the
Leibnitz identity $\{fg,h\} = f\{g,h\} + \{f,h\}g$. If Jacobi identity is
not required one speaks about pre-Poisson brackets. A mapping $F:M_1 \to
M_2$ of two manifolds equipped with Poisson brackets $\{\cdot,\cdot\}_1$
and $\{\cdot,\cdot\}_2$, respectively, is called {\em Poisson} if

\begin{displaymath}
\{f\circ F,g \circ F\}_1 = \{f,g\}_2 \circ F.
\end{displaymath}
In other words, a canonical mapping $F^*: C^\infty(M_2) \to C^\infty(M_1)$
is a Lie algebra homomorphism.

Let $M$ be a smooth manifold, and $*$ be a multiplication, i.e. a mapping
$M \times M \to M$. This immediately gives rise to a co-multiplication
(diagonal) $\Delta: C(M) \to C(M) \otimes C(M)$ on the algebra $C(M)$ of
functions on $M$ (with the pointwise multiplication). Namely,
\begin{displaymath}
\Delta(f)(x,y) = f(x * y)
\end{displaymath}
where $C(M \times M)$ is identified with $C(M) \otimes C(M)$.  As it is
well-known, algebra $C(M)$ is responsible for the topological structure of
$M$ while a diagonal $\Delta$ reflects a multiplication structure.

A Poisson bracket is said to be compatible with this multiplication  if
$* \colon M \times M \to M$ is a Poisson mapping where $M \times M$ is
equipped with a product Poisson bracket. In other words, the following
identity should be satisfied:
\begin{equation}\label{DiagComp}
\Delta(\{f,g\}) = \{\Delta(f),\Delta(g)\}
\end{equation}
where
\begin{displaymath}
\{p \otimes q, r \otimes s\} \bydef \{p,r\} \otimes qs + pr \otimes \{q,s\}.
\end{displaymath}
In particular, a bracket on a vector space compatible with addition
structure is exactly a Berezin--Lie one, i.e. a linear bracket (speaking about
linear, quadratic, etc., brackets we will always mean that a bracket of two
{\em linear} functions is linear, quadratic, etc., respectively).

The next-simplest case is quadratic Poisson brackets (see \cite{Skl}).
Such a bracket may be compatible only with a bilinear operation on the
vector space, i.e. with an algebra structure. One of the most important
cases, that of a full matrix algebra $\Mat(n,K)$ where $K = \Real$ or
$\Complex$, was investigated in detail in a series of works by
B~Kupershmidt \cite{Ku1,Ku2,Ku3}. He also studied conditions under which a
determinant (regarded as a function on $\Mat(n,K)$) is central.

In this paper a general description of Poisson brackets compatible with a
given algebra structure is given. A case of quaternion algebra is given a
special consideration. All the compatible brackets are enumerated, and a
explicitly described are brackets for which a norm is central.

Quantization of Poisson brackets compatible with algebra structures, as well
as their connection with quantum group theory is a subject of a forthcoming
paper.

\subsection{Quadratic brackets and differentiations}
Consider a vector space $A$ with a basis $\{e_i\}$, and let $x^i$ be
coordinate functions. A quadratic Poisson bracket is given by
\begin{equation}\label{DefPoi}
\{x^i,x^j\} = c_{kl}^{ij}x^kx^l,
\end{equation}
a summation over repeated indices will be always assumed. Symbol
$c_{kl}^{ij}$ is skew-symmetric with respect to upper indices, but
symmetry with respect to $k$ and $l$ is not generally assumed. Our task is
to study brackets (\ref{DefPoi}) compatible with the algebra structure in
$A$ given by the numerical structure constants $a_{kl}^i$, i.e.
\begin{equation}\label{DefAlg}
e_k \cdot e_l = a_{kl}^i e_i.
\end{equation}

With $A$ being an algebra, a tensor square $A \otimes A$ can also be given
an algebra structure by a componentwise multiplication. It is easy to see
that the set of symmetric tensors $\Symm(A \otimes A) \subset A \otimes A$
is then a subalgebra, while the set of skew-symmetric tensors $\Skew(A
\otimes A) \subset A \otimes A$ is a bimodule over $\Symm(A \otimes A)$. A
linear mapping $D: B \to V$ from algebra $B$ to a $B$-bimodule $V$ is
called a {\em differentiation} if it obeys the condition
\begin{equation}\label{DefDif}
D(p \cdot q) = p D(q) + D(p) q
\end{equation}
for all $p,\,q \in B$.
\begin{example}\label{IntDif}
Let $r \in V$, and the algebra $A$ is associative. Then the mapping
\begin{displaymath}
D_r(a) = ar - ra
\end{displaymath}
is a differentiation.
\end{example}

Constants $c_{kl}^{ij}$ from (\ref{DefPoi}) define also an operator $C:
\Symm(A \otimes A) \to \Skew(A \otimes A)$ by the formula
\begin{equation}\label{DefC}
C(e_k \otimes e_l + e_l \otimes e_k) = (c_{kl}^{ij} + c_{lk}^{ij}) e_i
\otimes e_j
\end{equation}

\begin{theorem}\label{CondComp}
The Poisson bracket (\ref{DefPoi}) is compatible with the multiplication
(\ref{DefAlg}) if and only if the operator $C$ given by (\ref{DefC}) is a
differentiation.
\end{theorem}

\begin{proof}
A co-multiplication $\Delta$ on the function algebra $C^\infty(A)$ is given
by
\begin{displaymath}
\Delta(x^i) = a_{kl}^i x_k \otimes x_k.
\end{displaymath}

The left-hand side of (\ref{DiagComp}) equals $c_{kl}^{ij} a_{ef}^k a_{gh}^l
x^e x^g \otimes x^f x^h$, so the coefficient at the term $x^m x^n \otimes
x^t x^y$ is
\begin{equation}\label{Coeff1L}
c_{kl}^{ij} (a^k_{mt} a^l_{ny} + a^k_{nt} a^l_{my} + a^k_{my} a^l_{nt} +
a^k_{ny} a^l_{mt})
\end{equation}
provided $m \ne n,\, t \ne y$, with obvious simplifications if some indices
are the same.

The right-hand side of (\ref{DiagComp}) equals $a_{pq}^i a_{rs}^j (x^p x^r
\otimes c_{wz}^{qs} x^w x^z + c_{uv}^{pr} x^u x^v \otimes x^q x^s)$, so the
coefficient at the term $x^m x^n \otimes x^t x^y$ is
\begin{equation}\label{Coeff1R}
(c_{mn}^{pr} + c_{nm}^{pr})(a^i_{pt}a^j_{ry} + a^i_{py}a^j_{rt}) +
(c_{ty}^{qs} + c_{yt}^{qs})(a^i_{mq}a^j_{ns} + a^i_{nq}a^j_{ms}),
\end{equation}
again, provided $m \ne n,\, t \ne y$. Compatibility condition means
coincidence of (\ref{Coeff1L}) and (\ref{Coeff1R}) for all
$i,\,j,\,m,\,n,\,t$, and $y$.

Write now differentiation identity (\ref{DefDif}) for the operator $C$ given
by (\ref{DefC}) and $p = e_m \otimes e_n + e_n \otimes e_m$ and $q = e_t
\otimes e_y + e_y \otimes e_t$. We suppose again that $m \ne n,\, t \ne y$,
with the formulas being obviously simplified in case some equality holds.
Then the coefficient at $e_k \otimes e_l$ in the left-hand side of
(\ref{DefDif}) equals
\begin{equation}\label{Coeff2L}
c_{kl}^{ij} (a^k_{mt} a^l_{ny} + a^k_{nt} a^l_{my} + a^k_{my} a^l_{nt} +
a^k_{ny} a^l_{mt}),
\end{equation}
and in the right-hand side,
\begin{equation}\label{Coeff2R}
(c_{mn}^{pr} + c_{nm}^{pr})(a^i_{pt}a^j_{ry} + a^i_{py}a^j_{rt}) +
(c_{ty}^{qs} + c_{yt}^{qs})(a^i_{mq}a^j_{ns} + a^i_{nq}a^j_{ms}).
\end{equation}
Coincidence of (\ref{Coeff1L}) with (\ref{Coeff2L}), and of (\ref{Coeff1R})
with (\ref{Coeff2R}), is easily observed.
\end{proof}

\begin{Remark}
All the above consideration do not make use of Jacobi identity, and
therefore apply for general pre-Poisson brackets as well.
\end{Remark}

\begin{example}\cite{FRT}
Consider a Poisson bracket
\begin{displaymath}
\{x^i,x^j\} = x^ix^j \quad \mbox{for $i < j$}.
\end{displaymath}
It is compatible with the co-multiplication
\begin{displaymath}
\Delta(x^i) = x^1 \otimes x^i
\end{displaymath}
i.e. with the structure of the ``first column algebra'' (algebra of $n
\times n$-matrices having nonzero elements only in the first column). The
corresponding differentiation is however not given by a commutator described
in Example \ref{IntDif}.
\end{example}

\subsection{Coboundary brackets}
Hereafter all the algebras are assumed associative. In the previous Section
any quadratic pre-Poisson bracket compatible with the algebra $A$ was
identified with a differentiation $C: \Symm(A \otimes A) \to \Skew(A
\otimes A)$. We restrict now our considerations to the {\em coboundary}
case, when this differentiation is {\em internal}, i.e. given by
\begin{equation}\label{IntDif2}
C(a) = ra - ar
\end{equation}
where $r \in A \otimes A$. In some cases this however exhausts all the
possible differentiations, e.g. if $A = \Mat(n)$ is a full matrix algebra.
For this algebra $r$ is necessarily a skew-symmetric tensor.

An important function on $A = \Mat(n)$ is a determinant $\det(M)$.
B~Kupershmidt in \cite{Ku3} gives a necessary and sufficient condition on
$r \in A \otimes A$ for $\det(M)$ to be a central function. It is
\begin{displaymath}
r_{ij}^{ik} = 0 \quad \mbox{for all $j,\,k$.}
\end{displaymath}

As a natural generalization of it, consider an algebra $A$ and a function
$det(M)$, a determinant in its left regular representation.

\begin{theorem}\label{DetCentr}
Let $r \in A \otimes A$ be as in (\ref{IntDif2}), and $r_1,\,r_2 \in A$ be
its traces (in a regular representation) in the first and the second
component, respectively. Then $det$ is a central function, if and only if
for all $x \in A$ the element $y = x r_1 - r_2 x$ is a left annulator of
$A$ (i.e.  $yz = 0$ for all $z \in A$).
\end{theorem}

The proof of Theorem \ref{DetCentr} simply copies Kupershmidt's
computations in \cite{Ku3}.

\subsection{Algebra of quaternions}
Consider coboundary pre-Poisson structures compatible with the algebra $H$
of quaternions. A condition that a commutator (\ref{IntDif2}) maps $\Symm(H
\otimes H)$ into $\Skew(H \otimes H)$ gives that the element $r$ can always
be chosen skew-symmetric. Let $\lnorm \cdot \rnorm$ be a standard norm in
$H$, and $x_1,\,\dots,\,x_4$ be coordinate functions with respect to a
standard basis $e_1 = {\bf 1},\,e_2 = {\bf i},\,e_3 = {\bf j},\,e_4 = {\bf
k}$.

\begin{theorem}\label{NormCent}
All the coboundary pre-Poisson brackets compatible with a multiplication in
$H$ such that $\lnorm \cdot \rnorm$ is a central function are given by the
following three-parameter family:
\begin{eqnarray}
\{x^1,x^2\} &=& x^2(bx^3 + ax^4) - c((x^3)^2 + (x^4)^2) \label{RowFirst}\\
\{x^1,x^3\} &=& x^3(ax^4 + cx^2) - b((x^4)^2 + (x^2)^2)\\
\{x^1,x^4\} &=& x^4(cx^2 + bx^3) - a((x^2)^2 + (x^3)^2)\\
\{x^2,x^3\} &=& -x^1(bx^2 + cx^3)\\
\{x^3,x^4\} &=& -x^1(ax^3 + bx^4)\\
\{x^4,x^2\} &=& -x^1(cx^4 + ax^2) \label{RowLast}
\end{eqnarray}
Moreover, all these brackets satisfy Jacobi identity (i.e. are not merely
pre-Poisson, but genuine Poisson brackets).
\end{theorem}

Theorem \ref{NormCent} is proved by straightforward (however cumbersome)
computations.

Since the norm is a central function with respect to brackets
(\ref{RowFirst})--(\ref{RowLast}), its level surfaces are Poisson
submanifolds of $H$. A surface $\lnorm x\rnorm =1$ is a group (inheriting
its multiplication from $H$) isomorphic to $\SU(2)$. Thus, Theorem
\ref{NormCent} gives us a three-parameter family of Poisson brackets in
$\SU(2)$ compatible with its Lie group structure, i.e., a three-parameter
family of Poisson Lie groups.

It would be very interesting to study similar questions for the algebra
${\bf O}$ of octaves. Kupershmidt in \cite{Ku3} studies connections
between pre-Poisson brackets on an algebra $A$ and on the group $\Aut(A)$
of its automorphisms. The thing is that the connected component of
$\Aut({\bf O})$ is isomorphic to the exceptional compact simple Lie group
$G_2$ which is currently a subject of intensive investigation.

The authors are grateful to Prof. B~Kupershmidt for sending his preprints.

\end{document}